\documentclass[preprint,prd,nofootinbib,floats,eqsecnum,tightenlines,showpacs,11pt]{revtex4-1}

\usepackage{graphicx}
\usepackage{setspace}
\usepackage{amsmath,amssymb,amsfonts,amsthm,latexsym}
\usepackage{enumerate}
\usepackage{colordvi}

\def\be{\begin{equation}}
\def\ee{\end{equation}}
\def\ba{\begin{eqnarray}}
\def\ea{\end{eqnarray}}
\def\sgn{\mathrm{sgn}}

\newcommand{\lp}{l_\text{Pl}}
\newcommand{\mb}{\bar{\mu}}

\begin{document}
%\preprint{\vbox{\baselineskip=12pt\rightline{APC-02/10-}\rightline{gr-qc/number}}}
\title{Unimodular Loop Quantum Cosmology}
\author{Dah-Wei Chiou}
\email{chiou@gravity.psu.edu}
\affiliation{Department of Physics, Beijing Normal University, Beijing 100875, China}

\author{Marc Geiller}
\email{mgeiller@apc.univ-paris7.fr}
\affiliation{APC - Astroparticule et Cosmologie\hbox{,} Universit\'e Paris Diderot Paris 7, Paris, France\bigskip\bigskip}

\begin{abstract}
\medskip
Unimodular gravity is based on a modification of the usual Einstein-Hilbert action that allows one to recover general relativity with a dynamical cosmological constant. It also has the interesting property of providing, as the momentum conjugate to the cosmological constant, an emergent clock variable. In this paper we investigate the cosmological reduction of unimodular gravity, and its quantization within the framework of flat homogeneous and isotropic  loop quantum cosmology. It is shown that the unimodular clock can be used to construct the physical state space, and that the fundamental features of the previous models featuring scalar field clocks are reproduced. In particular, the classical singularity is replaced by a quantum bounce, which takes place in the same condition as obtained previously. We also find that requirement of semiclassicality demands the expectation value of the cosmological constant to be small (in Planck units). The relation to spin foam models is also studied, and we show that the use of the unimodular time variable leads to a unique vertex expansion.
\end{abstract}


\maketitle

\section{Introduction}
Loop quantum gravity \cite{perez,rovelli-book,thiemann-book} (LQG) is a nonperturbative and background independent quantization of general relativity in which the geometry of space-time is treated quantum mechanically from the very beginning. The philosophy of LQG is based on taking seriously the lessons that we have learned from quantum mechanics and general relativity. It relies in particular on background independence \cite{ashtekar-background,ashtekar-background2,rovelli-background,smolin-background} and on the connection-dynamics formulation of general relativity \cite{ashtekar-canon,ashtekar-canon2}. The application of this theory of quantum gravity to symmetry reduced models of general relativity is known as loop quantum cosmology \cite{ashtekar-lqc,bojowald,vandersloot} (LQC). In the case of flat $\Lambda=0$ Friedmann-Lema\^itre-Robertson-Walker (FLRW) cosmologies with a massless scalar field, the theory strongly departs from general relativity in the Planck regime, where quantum gravity effects of LQG become significant and give rise to a new repulsive force \cite{aps}. When the matter energy density reaches the critical density in the deep Planck regime, this force is so strong that the universe undergoes a quantum bounce and the big bang singularity is avoided \cite{ashtekar-singularity}. It has been shown recently that this result is quite robust \cite{ashtekar-c-s}, and holds in $\Lambda\neq0$ cosmologies \cite{bentivegna}, $k=1$ closed cosmologies \cite{apsv,szulc}, $k=-1$ open cosmologies \cite{vandersloot-open}, and flat models with an inflationary potential \cite{aps-inflation}. The big bang singularity is also replaced in anisotropic Bianchi type I \cite{chiou-bianchiI,bianchiI}, type II \cite{bianchiII}, and type IX models \cite{bianchiIX}, as well as in Gowdy models \cite{gowdy}, in which there are infinitely many degrees of freedom.

In LQC with a massless scalar field $\phi$, the imposition of the quantum Hamiltonian constraint on states $\Psi$ leads to the evolution equation $\partial^2_\phi\Psi=-\Theta\Psi$, where $\Theta$ is a second order difference operator [unlike in Wheeler-DeWitt (WDW) theory, where it is a differential operator]. There are then two ways of constructing the physical state space. The first one is to focus on the constraint $\mathcal{C}\equiv\partial^2_\phi+\Theta=0$ as a whole and use the group averaging procedure \cite{marolf1,marolf2,marolf3} to map the kinematical states onto the physical ones. Alternatively, it is possible to see the scalar field $\phi$ as an internal time \textit{\`a la} Leibniz in which the evolution of physical quantities are computed in a fully relational manner \cite{gambini2}, and build a complete set of Dirac observables. However since scalar fields appear in the matter Hamiltonian through their momentum squared $p_\phi^2$, some difficulties can arise if we consider more complicated matter sectors (e.g. with a potential or phenomenological matter), since one needs to take the operator square root of the evolution equation in order to see it as a Schr\"odinger equation evolving in internal time $\phi$. In the present work we would like to study the possibility of using clocks that are not built with matter fields. We hope that this will help to study the robustness of LQC with alternative clock variables, and also shed light on some conceptual issues such as the problem of time \cite{ashtekar-stachel}.

All the theories of quantum gravity and quantum cosmology share the same deep philosophical questions, the most intriguing and exciting one being certainly the problem of time \cite{isham-butterfield}. By this, we refer in a very broad sense to the incompatibilities between the roles played by time in quantum mechanics and in general relativity. The notion of time we use in conventional quantum theory is deeply grounded in the concept of Newtonian physics, where there is an absolute time parameter external to any system, but this conception is simply incompatible with the formulation of diffeomorphism-invariant theories. When trying to address the problem of time, it is convenient to distinguish among three approaches: those in which a time variable is identified before the theory is quantized, those in which a time variable is identified at the quantum level, and finally those focusing on the timeless interpretation of quantum gravity. We refer the reader to \cite{isham} for a complete review. In this paper we rather focus on an approach known as unimodular gravity, in which time emerges at the classical level.

The idea of unimodular gravity can be traced back to Einstein, who noticed that imposing that the so-called unimodular condition $\det(g_{\mu\nu})=-1$ in the action principle leads to a theory of gravity in which the cosmological constant remains unspecified \cite{einstein1,einstein2}. It has therefore been argued that this theory might be used to solve the cosmological constant problems \cite{ng1,ng2,smolin-cc,weinberg}. The canonical formulation of unimodular gravity was studied by Henneaux and Teitelboim \cite{henneaux}, and by Unruh \cite{unruh-unimodular}. They have shown in particular how the cosmological constant arises as a dynamical variable accompanied by a canonically conjugate momentum corresponding to a time variable. Sorkin has also tried to implement this idea in a path integral approach to quantum gravity \cite{sorkin}. In the generally covariant formulation of Henneaux and Teitelboim, the cosmological constant and its conjugate time variable arise through the process of parametrizing \cite{hartle} a theory with a nonvanishing Hamiltonian. It is this point of view that we are going to adopt in order to investigate how the emergent time variable can be used as an alternative time variable in LQC. General issues concerning the use of the unimodular clock at the quantum level have been discussed by Kucha\v{r} \cite{kuchar-unimodular}, and Unruh and Wald \cite{unruh-wald}.

The aim of this paper is to investigate whether the predictions of LQC can be reproduced when a clock variable other than a scalar field is used to construct the physical sector and compute the evolution of observables. Additionally, we would like to see how the choice of the clock variable influences the sum over histories framework that reproduces in spirit the vertex expansion of spin foam models. In section \ref{section2}, we briefly review the covariant formulation of unimodular gravity and introduce the unimodular time variable. In section \ref{section3}, we investigate its cosmological reduction in the connection-triad variables and construct the classical theory with different matter sectors. Section \ref{section4} and \ref{section5} are devoted to reviewing the quantization procedure of LQC, and the usual construction of the physical Hilbert space with a scalar field clock. In section \ref{section6}, we use the unimodular clock variable to study the quantum dynamics of simplified LQC in the pure (without matter) de Sitter case. The issues of singularity resolution and occurrence of the quantum bounce are addressed in section \ref{section7}, where the theory with a nontrivial matter sector is studied. The relation with the vertex expansion of spin foam models is discussed in section \ref{section8}. We conclude and suggest some further developments in section \ref{section9}.

\section{The unimodular time coordinate}\label{section2}
We take as our starting point the generally covariant unimodular action of Henneaux and Teitelboim given by \cite{henneaux}
\be\label{HTaction}
S=\frac{1}{2\kappa}\int_\mathcal{M}d^4x\,\bigg[\sqrt{-g}\left(R-2\Lambda+\mathcal{L}_m\right)+2\Lambda\partial_\mu\tau^\mu\bigg].
\ee
Here $\kappa=8\pi G$, $\mathcal{L}_m$ is the Lagrangian for the matter sources, $\Lambda$ a scalar field that serves as a Lagrange multiplier, and $\tau^\mu=(\tau^0,\tau^a)$ a space-time vector density. Now consider a fixed coordinate system called unimodular coordinates, in which the volume four-form is independent  of the four metric $g_{\mu\nu}$, along with a second coordinate system in which the volume four-form $d^4x\,\sqrt{-g}$ appearing in the action (\ref{HTaction}) is calculated. It has been shown by Kucha\v{r} \cite{kuchar-unimodular} that the divergence $\partial_\mu\tau^\mu$ of the space-time vector density is simply the Jacobian that transforms from the general coordinate system to the unimodular coordinates. This formulation ensures that the action is invariant under the full group of diffeomorphisms.

Let us look at the equations of motion derived from the action (\ref{HTaction}). Performing its variation with respect to the multiplier $\Lambda$ gives the unimodular condition
\be\label{unimodular}
\sqrt{-g}=\partial_\mu\tau^\mu.
\ee
Varying $\tau^\mu$ leads to $\partial_\mu\Lambda=0$, which indicates that $\Lambda$ is a space-time constant that can be identified with the cosmological constant. Issues about the sign of $\Lambda$ will be discussed in the next section. Finally, variations with respect to the metric yield Einstein's field equations with an arbitrary cosmological constant.

The interpretation of the vector density $\tau^\mu$ is the following. Let us suppose that the manifold $\mathcal{M}$ is diffeomorphic to $\Sigma\times\mathbb{R}$, where $\Sigma$ is a compact three-manifold. We can integrate (\ref{unimodular}) over the space-time region $\mathcal{R}\subset\mathcal{M}$ bounded by two spacelike surfaces $\Sigma(t_1)$ and $\Sigma(t_2)$, with the parameter $t\in\mathbb{R}$ labeling the foliation of $\mathcal{M}$ into Cauchy surfaces. The upshot is
\be
\int_\mathcal{R}d^4x\,\partial_\mu\tau^\mu=\int_\mathcal{R}d^4x\,\sqrt{-g}=\text{four volume enclosed between $\Sigma(t_1)$ and $\Sigma(t_2)$}.
\ee
As we will see in what follows, the cosmological constant $\Lambda$ may be regarded as the momentum conjugate to a dynamical variable which may be interpreted as a cosmological time parameter associated to a surface $\Sigma(t)$. This time coordinate,
\be\label{T}
T(t)=\int_{\Sigma(t)}d^3x\,\tau^0,
\ee
is just the four volume preceding the spacelike surface $\Sigma(t)$, a function that increases continuously along any future directed timelike curve. This shows that the evolution of the vector density $\tau^\mu$ is a pure gauge, except for the mode zero $\tau^0$ which satisfies
\be
T(t_2)-T(t_1)=\text{four volume enclosed  between $\Sigma(t_1)$ and $\Sigma(t_2)$}.
\ee
The unimodular time coordinate acts in fact as a label of equivalence classes of spacelike hypersurfaces with zero four volume separation, which explains its non local nature.

\section{Cosmological reduction}\label{section3}
In this section, we introduce the $3+1$ description of the unimodular theory of gravity in connection-triad variables, and then reduce it to flat homogeneous and isotropic cosmological models. We then derive the classical equations of motion with and without matter sectors and discuss the presence of singularities.

\subsection{Hamiltonian framework}
Following the formulation of LQG in terms of the connection-triad variables, we introduce  a triad $e^a_i$ and its dual cotriad $\omega^i_a$, such that $e^a_i\omega^j_a=\delta^j_i$. Here Latin letters from the middle of the alphabet are Lie algebra indices, Latin letters from the beginning of the alphabet are spatial indices, and both are ranging from 1 to 3. The three metric $q_{ab}$ on the spacelike hypersurface $\Sigma$ is related to the cotriad via
\be
q_{ab}=\omega^i_a\omega^j_b\delta_{ij}.
\ee
The spatial geometry is encoded in the densitized triad
\be
E^a_i\equiv\frac{1}{2}\epsilon^{abc}\epsilon_{ijk}\omega^j_b\omega^k_c,
\ee
which serves as our basic phase space variable along with the $su(2)$-valued Ashtekar-Barbero connection
\be
A^i_a=\Gamma^i_a+\gamma K^i_a,
\ee
where $\Gamma^i_a$ is the spin connection, $K^i_a$ the extrinsic curvature one-form, and $\gamma$ the Barbero-Immirzi parameter. With these new variables, the action (\ref{HTaction}) can be recast in the $3+1$ form as \cite{bombelli}
\be\label{CanAction}
S=\int dt\int_\Sigma d^3x\,\left[\frac{1}{\kappa\gamma}E^a_i\dot{A}^i_a-N\mathcal{H}-N^a\mathcal{H}_a-N^i\mathcal{G}_i
-\frac{1}{\kappa}\left(N\Lambda\sqrt{q}-\Lambda\dot{\tau}^0+\tau^a\partial_a\Lambda\right)\right],
\ee
where $N$ is the lapse function enforcing the scalar constraint
\be\label{scalar}
\mathcal{H}=\frac{E^a_iE^b_j}{2\kappa\sqrt{q}}\left(\epsilon^{ij}_{~~k}F^k_{ab}-2\left(1+\gamma^2\right)K^i_{[a}K^j_{b]}\right),
\ee
where $F^k_{ab}=\partial_aA^k_b-\partial_bA^k_a+\epsilon^{~~k}_{ij}A^j_aA^k_b$ is the curvature of the connection $A^i_a$, and $2K^i_{[a}K^j_{b]}=K^i_aK^j_b-K^i_bK^j_a$.
$N^a$ is the shift vector multiplying the vector (or diffeomorphism) constraint $\mathcal{H}_a$, and $N^i$ a multiplier used to enforce the Gauss constraint $\mathcal{G}_i$. We see from this action that $\tau^a$ acts as a multiplier enforcing the first class constraint $\mathcal{G}_a\equiv\partial_a\Lambda=0$. The symplectic structure is given by
\be
\lbrace A^i_a(x),E^b_j(y)\rbrace=\kappa\gamma\delta^b_a\delta^i_j\delta^3(x,y),
\qquad
\lbrace\tau^0(x),\Lambda(y)\rbrace=\kappa\delta^3(x,y).
\ee
The total Hamiltonian is
\be
H=\int_\Sigma d^3x\,\left[N\left(\mathcal{H}+\mathcal{H}_\Lambda\right)+N^a\mathcal{H}_a+N^i\mathcal{G}_i+\frac{1}{\kappa}\tau^a\mathcal{G}_a\right],
\ee
where the new term in the scalar constraint is given by
\be
\mathcal{H}_\Lambda=\frac{1}{\kappa}\Lambda\sqrt{q}.
\ee
It enables us to compute the evolution of our canonical variables. In particular, we get $\dot{\Lambda}=0$ and
\be
\dot{\tau}^0=\kappa\frac{\delta H}{\delta\Lambda}=N\sqrt{q}=\sqrt{-g},
\ee
which agrees with our definition of the cosmological time coordinate $T(t)$ given in (\ref{T}).

Now we investigate the symmetry reduced model obtained by restricting ourselves to spatially flat, homogeneous and isotropic cosmologies. The spatial manifold $\Sigma$ is topologically $\mathbb{R}^3$, and since it is not compact, we restrict all integrations to a fixed fiducial cell $\mathcal{V}$ of fiducial volume $V_0$.  $\Sigma$ is endowed with a fiducial background flat metric
\be
{}^o{q}_{ab}={}^o\omega^i_a\,{}^o\omega^j_b\delta_{ij},
\ee
where ${}^o\omega^i_a$ is a fixed cotriad. Isotropic connections and densitized triads are characterized by
\begin{subequations}
\ba
\label{gauge}
A^i_a&=&V_0^{-1/3}c(t)\,{}^o\omega^i_a,\\
E^a_i&=&V_0^{-2/3}\sqrt{{}^oq}p(t)\,{}^oe^a_i.
\ea
\end{subequations}
The three metric is related to the fiducial metric by
\be
q_{ab}=a^2\left({}^o{q}_{ab}\right)=a^2\left({}^o\omega^i_a\,{}^o\omega^j_b\delta_{ij}\right),
\ee
where $a(t)$ is the scale factor, which satisfies $|p|=a^2V_0^{2/3}$. We also have the relation $\sqrt{q}=\sqrt{{}^oq}|p|^{3/2}$. Because we are using the fiducial cell $\mathcal{V}$, our phase space coordinates $c$ and $p$ are left unchanged if we rescale the fiducial metric ${}^oq_{ab}$. Notice that $p$ can be either positive or negative, which corresponds to two orientations of the triad. Since the metric is left unchanged under the transformation $e^a_i\rightarrow-e^a_i$ and we do not consider fermions in our model, the physics will also remain unchanged. Thus this orientation reversal of the triad represents a gauge transformation that needs to be handled.

Because the model that we describe is spatially flat, the spin connection $\Gamma^i_a$ vanishes and thus $A^i_a=\gamma K^i_a$. Since derivatives of $K^i_a$ do also vanish, we find that
\be
2K^i_{[a}K^j_{b]}=\frac{1}{\gamma^2}\epsilon^{ij}_{~~k}F^k_{ab},
\ee
and the gravitational part of the Hamiltonian can be expressed as
\be\label{LQCham}
\mathcal{H}=-\frac{E^a_iE^b_j}{2\kappa\gamma^2\sqrt{q}}\epsilon^{ij}_{~~k}F^k_{ab}.
\ee
Having fixed the gauge and diffeomorphism freedom, the constraints $\mathcal{H}_a$, $\mathcal{G}_i$ and $\mathcal{G}_a$ are automatically satisfied. Thus we are left with the sole scalar constraint, and one can show that the $3+1$ action (\ref{CanAction}) rewrites
\be
S=\int dt\left(\frac{3}{\kappa\gamma}p\dot{c}
+\frac{1}{\kappa}\Lambda\dot{T}
+N\frac{3}{\kappa\gamma^2}\sqrt{|p|}c^2
-\frac{1}{\kappa}N\Lambda|p|^{3/2}\right).
\ee
The canonical term gives us the Poisson brackets
\be
\lbrace c,p\rbrace=\frac{\kappa\gamma}{3},
\qquad
\lbrace T,\Lambda\rbrace=\kappa,
\ee
and the Hamiltonian is
\be\label{hamiltonian}
H=\int_\mathcal{V} d^3x\,N\left(-\frac{3}{\kappa\gamma^2}\sqrt{|p|}c^2+\frac{1}{\kappa}\Lambda|p|^{3/2}\right)
\equiv\int_\mathcal{V} d^3x\,N\left(\mathcal{H}+\mathcal{H}_\Lambda\right).
\ee
We see that the first term is the standard symmetry reduced gravitational part of the Hamiltonian constraint, whereas the second one corresponds to matter sourced by a cosmological constant $\Lambda$. The interesting fact is precisely that this cosmological constant has the time variable $T$ for conjugate momentum, which will enable us to obtain an equation evolving in time $T$ when imposing the quantum Hamiltonian constraint on wave functions.

\subsection{Classical equations of motion}
Given the Hamiltonian (\ref{hamiltonian}), we derive the classical equations of motion for our basic variables. It is obvious that $\Lambda$ is a constant of motion since its conjugate momentum $T$ does not appear in the Hamiltonian. With a choice of lapse $N=1$, the equation of motion for $T$ is given by
\be\label{T dot}
\dot{T}=\lbrace T,\mathcal{H}+\mathcal{H}_\Lambda\rbrace=
\kappa\frac{\partial\mathcal{H}_\Lambda}{\partial \Lambda}=|p|^{3/2},
\ee
and that for $p$ is given by
\be\label{p dot}
\dot{p}=\lbrace p,\mathcal{H}+\mathcal{H}_\Lambda\rbrace
=-\frac{\kappa\gamma}{3}\frac{\partial\mathcal{H}}{\partial c}
=\frac{2}{\gamma}\sqrt{|p|}\,c,
\ee
where the dot denotes derivation with respect to proper time $t$ (the choice $N=a=\sqrt{|p|}$ would correspond to conformal time). The constraint that the Hamiltonian is vanishing gives us
\be\label{c}
c=\pm\gamma\sqrt{\frac{\Lambda|p|}{3}},
\ee
where the plus and minus signs ($\pm$) correspond to the expanding and contracting solutions respectively as will be seen shortly. Note that, without the inclusion of matter, the Hamiltonian constraint demands $\Lambda$ to be positive\footnote{\label{footnote on ADS}In the anti-de Sitter universe, the space is not globally hyperbolic and we do not have a well-posed initial value problem by specifying the initial information on a spacelike slice \cite{hawking-ellis,wald}. Therefore, the Hamiltonian framework with the $3+1$ splitting is not suitable for the anti-de Sitter solution. On the other hand, with the inclusion of matters, both positive and negative values of $\Lambda$ are permitted.}.

Substituting (\ref{c}) into (\ref{p dot}) yields
\be\label{p dot 2}
\dot{p}=\pm2\sqrt{\frac{\Lambda}{3}}\ |p|,
\ee
which gives the solution
\be
|p|=|p_0|\exp\left(\pm2\sqrt{\frac{\Lambda}{3}}\ t\right)
\ee
as a function of the proper time $t$. As expected, at the classical level, the pure (without matter) de Sitter universe described here takes infinite proper time to reach $p=0$ or $p=\infty$ and does not exhibit any singularity. Indeed the matter density remains constant and proportional to $\Lambda$ and therefore does not blow up at any epoch of the evolution. The same thing happens to the Ricci curvature scalar since it is proportional to the cosmological constant. To create a singularity and reach it in finite proper time, we have to add to the theory some matter with nonconstant matter density.

On the other hand, combining (\ref{T dot}) and (\ref{p dot 2}) yields
\be
\dot{p}=\pm\frac{2\dot{T}}{\sqrt{|p|}}\sqrt{\frac{\Lambda}{3}},
\ee
which can be integrated to find the solution of $p$ in terms of $T$
\be\label{class}
|p(T)|^{3/2}-|p(T_0)|^{3/2}
=\pm\sqrt{3\Lambda}\,(T-T_0).
\ee
The triad $p$ is a monotonic function of $T$. The solution gives either an expanding or a contracting universe according to the plus or minus sign ($\pm$) respectively. Note that it takes finite time $T$ to reach $p=0$, but this does not correspond to a singularity as remarked above.

Following what is usually done in LQC, we can add a scalar field with Hamiltonian
\be
\mathcal{H}_\phi=\frac{1}{2}\frac{p_\phi^2}{|p|^{3/2}}+|p|^{3/2}U(\phi),
\ee
where $U(\phi)$ is a potential, and the Poisson bracket is $\lbrace \phi,p_\phi\rbrace=1$. To describe other matter sources, such as radiation, dust or any phenomenological component, it is convenient to consider the general constant equation of state $P_w=w\rho_w$. The continuity equation
\be
\dot{\rho}_w+3\frac{\dot{a}}{a}(P_w+\rho_w)=0
\ee
can be integrated to find the dilution law
\be
\rho_w=\beta a^{-3(1+w)}=\beta |p|^{-3(1+w)/2},
\ee
where $\beta$ is a constant factor. Dust has an equation of state $w=0$, radiation is described by $w=1/3$, a massless scalar field by $w=1$, and finally slow-roll inflation corresponds to $w=-1$ (as well as a cosmological constant term). The Hamiltonian describing this generic case is
\be
\mathcal{H}_w=a^3\rho_w=|p|^{3/2}\rho_w=\beta|p|^{-3w/2}.
\ee
The total matter density is given by
\be\label{total density}
\rho_{\Lambda+w}\equiv\rho_\Lambda+\rho_w
=\frac{\Lambda}{\kappa}+\beta|p|^{-3(1+w)/2},
\ee
which blows up at $p=0$ if $w>-1$ (ordinary matter) and at $p=\infty$ if $w<-1$ (exotic matter). In the following, we will show that $p=0$ or $p=\infty$ corresponds to the singularity for $w>-1$ and $w<-1$, respectively.

Since the total Hamiltonian $\mathcal{H}+\mathcal{H}_\Lambda+\mathcal{H}_w$ is vanishing as a constraint, we have
\be
c=\pm\gamma\sqrt{\frac{\Lambda|p|}{3}+\frac{\beta\kappa}{3}|p|^{-(3w+1)/2}}.
\ee
Note that in the presence of the matter sector, the Hamiltonian constraint admits both positive and negative values of $\Lambda$.
The equation of motion for $T$ is again $\dot{T}=|p|^{3/2}$, and for the triad we find
\begin{subequations}
\ba
\label{p dot with matter 1}
\dot{p}&=&\pm2\sqrt{\frac{\Lambda|p|^2}{3}+\frac{\beta\kappa}{3}|p|^{-(3w-1)/2}}\\
\label{p dot with matter 2}
&=&\pm2\dot{T}\sqrt{\frac{\Lambda}{3|p|}+\frac{\beta\kappa}{3}|p|^{-(3w+5)/2}}.
\ea
\end{subequations}
For $w>-1$ (respectively $w<-1$), when $p=0$ (respectively $p=\infty$) is approached, the second term inside the square root in (\ref{p dot with matter 1}) dominates the first term, and thus we have
\be
\dot{p}\approx\pm2\sqrt{\frac{\beta\kappa}{3}}\ |p|^{-(3w-1)/4},
\ee
which leads to
\be
|p(t)|^{3(w+1)/4}-|p(t_0)|^{3(w+1)/4}
\approx\frac{w+1}{2}\sqrt{3\beta\kappa}\left(t-t_0\right).
\ee
This shows that it takes finite proper time $t$ to reach $p=0$ for $w>-1$ and $p=\infty$ for $w<-1$. Therefore, with the inclusion of matter, the classical solution encounters the big bang or big crunch singularity for $w>-1$ and the big rip singularity for $w<-1$.

On the other hand, in the quantum theory (for both WDW and LQC), choosing $w=-2$ in particular makes the algebra much simpler. In order to compare the quantum evolution with the classical one, we consider $w=-2$, and (\ref{p dot with matter 2}) yields the solution\footnote{The solution shows that it takes infinite time $T$ to reach $p=\infty$. This nevertheless is a true big rip singularity since it is reached in finite proper time $t$.}
\be
\sqrt{\frac{\Lambda}{3}+\frac{\beta\kappa}{3}|p(T)|^{3/2}}
-\sqrt{\frac{\Lambda}{3}+\frac{\beta\kappa}{3}|p(T_0)|^{3/2}}
=\pm\frac{\beta\kappa}{2}\left(T-T_0\right),
\ee
where the plus and minus signs ($\pm$) correspond to the expanding and contracting solutions respectively. It follows that
\be\label{class with matter}
|p(T)|^{3/2}-|p(T_0)|^{3/2}
=\pm\sqrt{3\Lambda+3\beta\kappa\,|p(T)|^{3/2}}\left(T-T_0\right)
-\frac{3\beta\kappa}{4}\left(T-T_0\right)^2,
\ee
which will be used to compare with the counterparts of the quantum theory.

\section{Quantum kinematics}\label{section4}
In this section we are going to study the loop quantization of the unimodular cosmological theory at hand. For this we follow the usual Dirac procedure \cite{dirac}, in which the constraints are first quantized and then imposed on the wave functionals. Again, because of homogeneity and isotropy, the diffeomorphism and Gauss constraints are satisfied, as well as the constraint $\mathcal{G}_a$. We are left with the sole scalar Hamiltonian constraint.

As in the full theory, there is no operator corresponding the the connection itself \cite{thiemann-book}. The only well-defined operator is the one corresponding to the holonomy of the connection, and therefore we need to go back to the expression (\ref{LQCham}) of the gravitational Hamiltonian and express the curvature and the triad in terms of the holonomies. The holonomy of the connection $A^i_a$ along an oriented edge $\lambda\,{}^oe^a_i$ is given by \cite{alb}
\be
h^{(\lambda)}_i=\cos\left(\frac{\lambda c}{2}\right)\openone+2\sin\left(\frac{\lambda c}{2}\right)\tau_i,
\ee
where $\openone$ is the unit $2\times2$ matrix and $\tau_k$ a basis of the Lie algebra $su(2)$ such that $2i\tau_k=\sigma_k$, where $\sigma_k$ is a Pauli matrix. To impose discreteness in the spirit of the so-called improved dynamics scheme of LQC, we choose $\lambda$ to be a specific function $\mb(p)$ of the triad given by \cite{aps}
\be\label{mubar}
\mb^2|p|=\Delta\equiv2\sqrt{3}\pi\gamma\lp^2,
\ee
where $\lp^2=G\hbar$ is the Planck length squared, and $\lambda^2\equiv\Delta$ the area gap of LQG corresponding to the smallest eigenvalue of the surface operator. For the sake of convenience, we introduce the new canonical variables
\be
b\equiv\frac{c}{\sqrt{|p|}},
\qquad
\nu\equiv\sgn(p)\frac{|p|^{3/2}}{2\pi\gamma\lp^2},
\ee
which have canonical Poisson bracket
\be
\lbrace b,\nu\rbrace=\frac{2}{\hbar},
\ee
and satisfy $\mb c=\lambda b$.

Now we choose to work in the $\nu$ representation, where states are given by wave functions $\Psi(\nu)$. The kinematical Hilbert space is $\mathfrak{H}^{\text{kin}}=\mathfrak{H}^{\text{kin}}_{\text{grav}}\otimes\mathfrak{H}^{\text{kin}}_{\text{matt}}$. In the next three sections, we will study the different matter sectors in relation to the choice of clock variable. For the moment we focus on the gravitational sector. The corresponding kinematical Hilbert space is the space of square integrable functions on the Bohr compactification of the real line $\mathfrak{H}^{\text{kin}}_{\text{grav}}=L^2(\mathbb{R}_{\text{Bohr}},d\mu_{\text{Bohr}})$. Elements of this gravitational kinematical Hilbert space are wave functions $\Psi(\nu)$ with finite norm
\be
||\Psi||^2=\sum_\nu|\Psi(\nu)|^2,
\ee
and an orthonormal basis is given by $|\nu\rangle$ with
\be
\langle\nu|\nu'\rangle=\delta_{\nu\nu'}.
\ee
The operator measuring the volume of the cell $\mathcal{V}$ acts like
\be
\hat{V}\Psi(\nu)\equiv2\pi\gamma\lp^2|\hat{\nu}|\Psi(\nu)=2\pi\gamma\lp^2|\nu|\Psi(\nu),
\ee
and the holonomies are promoted to operators satisfying
\be
\widehat{e^{\pm i\mb c}}\Psi(\nu)\equiv\widehat{e^{\pm i\lambda b}}\Psi(\nu)=\Psi(\nu\pm2\lambda).
\ee
Our task is now to promote the Hamiltonian to a quantum operator. To do so, we need to go back to the full expression (\ref{LQCham}) and express the curvature $F^k_{ab}$ and the inverse triad in terms of the holonomies. As in the usual improved dynamics quantization of LQC, the action of the gravitational Hamiltonian can be expressed as
\be
\hat{\mathcal{H}}\Psi(\nu,T)=\widehat{\sin\left(\lambda b\right)}\hat{A}(\nu)\widehat{\sin\left(\lambda b\right)}\Psi(\nu,T),
\ee
where $\hat{A}$ is a self-adjoint and negative definite operator on $\mathfrak{H}^{\text{kin}}_{\text{grav}}$ defined by
\be
\hat{A}(\nu)\Psi(\nu)=-\frac{3\pi\lp^2}{\kappa\gamma\lambda^3}|\nu|\Big||\nu+\lambda|-|\nu-\lambda|\Big|\Psi(\nu).
\ee
In the matter part of the Hamiltonian, the inverse triad $|p|^{-3/2}$ appears and we need to quantize it properly. Thiemann's trick allows us to compute it in terms of holonomies, and its expression in the $\nu$ representation if found to be
\ba
\widehat{V^{-1}}\Psi(\nu)=\widehat{|p|^{-3/2}}\Psi(\nu)&=&\frac{27}{64\lambda^3}\Big||\nu+\lambda|^{2/3}-|\nu-\lambda|^{2/3}\Big|^3\Psi(\nu)\cr\cr
&\equiv&\hat{B}(\nu)\Psi(\nu).
\ea

Let us consider the general case in which a cosmological constant, a scalar field with potential $U(\phi)$, and some matter with equation of state $w$ are present in the matter sector. We just gave a prescription to write down the quantum operator corresponding to the gravitational part. For the matter part, we are going to use the usual Schr\"odinger representation, keeping in mind that in the unimodular theory, the cosmological constant is also turned into a quantum operator. Requiring that the total quantum Hamiltonian operator (for a generic matter sector) annihilates the wave functionals $\Psi\in\mathfrak{H}^{\text{kin}}$ leads to the general equation
\ba\label{full}
\bigg(\hat{\mathcal{H}}+\frac{1}{2\kappa}\hat{\Lambda}\hat{B}(\nu)^{-1}+\frac{1}{2}\hat{p}_\phi^2\hat{B}(\nu)+U(\hat{\phi})\hat{B}(\nu)^{-1}+\beta \hat{B}(\nu)^w\bigg)\Psi=0.
\ea
Now for the sake of simplicity, we shall adopt the two simplifications of the Hamiltonian constraint which render the model exactly soluble and simplifies the derivation of the WDW limit \cite{ashtekar-c-s}. This can be done by using the following simpler expressions for the operators $\hat{A}(\nu)$ and $\hat{B}(\nu)$:
\be
\hat{A}(\nu)=-\frac{6\pi\lp^2}{\kappa\gamma\lambda^2}|\nu|,
\qquad
\hat{B}(\nu)=\frac{1}{2\pi\gamma\lp^2|\nu|}.
\ee
The first simplification is an extremely good approximation, and it is exact if the wave function $\Psi(\nu)$ has support in the superselected sector $\nu=4n\sqrt{\Delta}\lp,\,n\in\mathbb{Z}$. The second approximation is a bit more restrictive, but it has been shown that the error decreases extremely rapidly for high values of $\nu/\lambda$ \cite{ashtekar-c-s}.

Finally, we require that physical states lie in one of the irreducible representations (either symmetric or antisymmetric) of the gauge transformation corresponding to the orientation reversal of the triad $\Pi:\nu\longmapsto\Pi(\nu)=-\nu$. Since we do not consider fermions in the present model, wave functions are assumed to be symmetric: $\Pi\Psi(\nu)\equiv\Psi(-\nu)=\Psi(\nu)$.

\section{Scalar field clock}\label{section5}
In this part we investigate the construction of the physical sector of LQC. To begin with, we forget about unimodular gravity and review the procedure that is usually followed when matter clocks such as scalar fields are used to describe evolution \cite{aps}.

In standard treatments of LQC, the matter sector corresponds to a free scalar field, which is monotonic on all classical solutions (this is also the case when $k=1$ and $\Lambda\neq0$), as can be seen from the classical Hamiltonian equations of motion. This suggests that the scalar field is well suited to be an internal clock with respect to which we can evolve the other physical variables. If the scalar field has a potential, we have $\dot{p}_\phi\neq0$, and its evolution does not remain monotonic. Thus we have to reset the scalar field clock at different instants of parameter time if we still want to use it as a clock. The free scalar field is quantized in the usual Schr\"odinger representation on $\mathfrak{H}^{\text{kin}}_{\phi}=L^2(\mathbb{R},d\phi)$ given by
\be
\hat{\phi}\Psi(\phi)=\phi\Psi(\phi),
\qquad
\hat{p}_\phi\Psi(\phi)=-i\hbar\frac{\partial\Psi(\phi)}{\partial\psi}.
\ee
If we choose a gauge in which the lapse $N=1$ and adopt the ordering used in \cite{ashtekar-c-s}, the action of the quantum Hamiltonian operator on wave functionals $\Psi(\nu,\phi)$ gives the evolution equation
\ba\label{LQCev}
\frac{\partial^2\Psi(\nu,\phi)}{\partial\phi^2}&=&-3\pi G\nu\frac{\widehat{\sin(\lambda b)}}{\lambda}\nu\frac{\widehat{\sin(\lambda b)}}{\lambda}\Psi(\nu,\phi)\cr\cr
&=&\frac{3\pi G}{4\lambda^2}\nu\big[(\nu+2\lambda)\Psi(\nu+4\lambda)-2\nu\Psi(\nu)+(\nu-2\lambda)\Psi(\nu-4\lambda)\big]\cr\cr
&\equiv&-\Theta\Psi(\nu,\phi).
\ea
$\Theta$ is a positive self-adjoint operator on $\mathfrak{H}^{\text{kin}}_{\text{grav}}=L^2(\mathbb{R}_{\text{Bohr}},B(\nu)d\mu_{\text{Bohr}})$. For each $\epsilon\in[0,4\lambda)$, there is a Hilbert space $\mathfrak{H}^{\text{kin}}_\epsilon\subset\mathfrak{H}^{\text{kin}}$ in which states have support on the lattice points $\nu=\epsilon+4n\lambda,\;n\in\mathbb{Z}$. Since each of these subspaces is preserved under the evolution, there is superselection, and in the remainder of this work we will focus on the $\epsilon=0$ lattice. The difference operator $\Theta$ acts only on the $\nu$ argument of $\Psi(\nu,\phi)$, and the form of the constraint equation indicates that it is natural to view $\phi$ as an emergent time. Elements $\Psi(\nu,\phi)\in\mathfrak{H}^{\text{kin}}=\mathfrak{H}^{\text{kin}}_{\text{grav}}\otimes\mathfrak{H}^{\text{kin}}_{\phi}$ are wave functions with finite kinematical norm
\be
||\Psi||^2=\int_{\mathbb{R}}d\phi\sum_\nu|\Psi(\nu,\phi)|^2,
\ee
and an orthonormal basis in $\mathfrak{H}^{\text{kin}}$ is given by $|\nu,\phi\rangle$ with
\be
\langle\nu,\phi|\nu',\phi'\rangle=\delta_{\nu\nu'}\delta(\phi,\phi').
\ee
Now we can build a set of complete Dirac observables, the scalar field momentum $p_\phi$ being a trivial one. At a fixed value $\phi=\phi_0$ of internal time, the volume $V|_{\phi_0}$ is also a Dirac observable. Since the space of positive and negative frequency solutions is left invariant under the action of these Dirac observables, there is superselection. As in Klein-Gordon theory, we therefore choose to work with either set. Then elements $\Psi(\nu,\phi)$ of the physical Hilbert space $\mathfrak{H}^{\text{phys}}$ are the positive frequency solutions to the quantum constraint (\ref{LQCev}), \textit{i.e.} solutions satisfying
\be\label{sqrtLQC}
-i\frac{\partial\Psi(\nu,\phi)}{\partial\phi}=\sqrt{\Theta}\Psi(\nu,\phi).
\ee
Notice that this equation is equivalent to (\ref{LQCev}) because the evolution operator $\Theta$ is independent of $\phi$. Indeed, if we act with the operator $\partial/\partial\phi$ on (\ref{sqrtLQC}), we obtain
\be
-i\frac{\partial^2\Psi(\nu,\phi)}{\partial\phi^2}=\frac{\partial}{\partial\phi}\big(\sqrt{\Theta}\big)\Psi(\nu,\phi)+i\Theta\Psi(\nu,\phi),
\ee
and recover (\ref{LQCev}) since the first term on the right-hand side does vanish.

The physical inner product is defined at a fixed instant of time as
\be\label{physicalSP with phi}
(\Psi_1,\Psi_2)_{\text{phys}}=\frac{\lambda}{\pi}\sum_{\nu=4n\lambda}\frac{1}{|\nu|}\bar{\Psi}_1(\nu,\phi_0)\Psi_2(\nu,\phi_0),
\ee
and is independent of the choice of $\phi_0$ since the operator $\Theta$ is self-adjoint and positive definite with respect to $(\cdot\,,\cdot)_{\text{phys}}$ given above. Note that the state with support at $\nu=0$ yields infinite norm and hence is not an element of the physical Hilbert space\footnote{It should be noted that, as emphasized in \cite{ashtekar-c-s}, the decoupling of the state $|\nu=0\rangle$ does not imply resolution of the singularity.}. Now we explicitly define the action of the set of Dirac observables. The scalar field momentum acts like
\be
\hat{p}_\phi\Psi(\nu,\phi)=-i\hbar\frac{\partial\Psi(\nu,\phi)}{\partial\phi}=\hbar\sqrt{\Theta}\Psi(\nu,\phi),
\ee
and the volume at fixed instant of time gives a family of relational observables defined as
\be
\hat{V}|_{\phi_0}\Psi(\nu,\phi)=2\pi\gamma\lp^2e^{i\sqrt{\Theta}(\phi-\phi_0)}|\hat{\nu}|\Psi(\nu,\phi_0)=2\pi\gamma\lp^2e^{i\sqrt{\Theta}(\phi-\phi_0)}|\nu|\Psi(\nu,\phi_0),
\ee
which represents the action of the volume operator on the state $\Psi(\nu,\phi)$ at $\phi=\phi_0$ and then its evolution from $\phi_0$ to $\phi$ with the equation (\ref{sqrtLQC}).

\section{Unimodular clock variable}\label{section6}
In this section we study the quantization of the pure (without matter) de Sitter cosmological model corresponding to the unimodular theory. We take the wave functionals to be of the form $\Psi(\nu,T)$, and the usual Schr\"odinger representation on $\mathfrak{H}^{\text{kin}}_{\Lambda}=L^2(\mathbb{R},dT)$ is used for the cosmological constant part:
\be
\hat{T}\Psi(\nu,T)=T\Psi(\nu,T),
\qquad
\hat{\Lambda}\Psi(\nu,T)=-\kappa i\hbar\frac{\partial\Psi(\nu,T)}{\partial T}.
\ee
Using again the simplified scheme introduced above, the total constraint $(\hat{\mathcal{H}}+\hat{\mathcal{H}}_\Lambda)\Psi=0$ takes the form
\ba
-i\frac{\partial\Psi(\nu,T)}{\partial T}&=&\frac{3}{\kappa\hbar\gamma^2}\frac{\widehat{\sin(\lambda b)}}{\lambda}\frac{\widehat{\sin(\lambda b)}}{\lambda}\Psi(\nu,T)\label{LQC}\\
&=&-\frac{3}{4\kappa\hbar\gamma^2\lambda^2}\big[\Psi(\nu+4\lambda)-2\Psi(\nu)+\Psi(\nu-4\lambda)\big]\nonumber\\
&\equiv&\tilde{\Theta}\Psi(\nu,T),\nonumber
\ea
where the difference operator $\tilde{\Theta}$ is again self-adjoint and positive definite on $\mathfrak{H}^{\text{kin}}_{\text{grav}}$. Before studying the LQC dynamics of this theory, we would like to first ignore the effects coming from quantum geometry and focus on its WDW limit. This will make more transparent the construction of the physical state space and the Dirac observables, and facilitate the comparison between the two quantizations.

\subsection{Wheeler-DeWitt limit}
The WDW limit is obtained from the LQC dynamical equation (\ref{LQC}) by taking the formal limit $\lambda\rightarrow0$. In the so-called $b$ representation, the constraint equation takes the following Schr\"odinger-like form:
\be\label{WDW}
-i\frac{\partial\Psi(b,T)}{\partial T}=\frac{3}{\kappa\hbar\gamma^2}b^2\Psi(b,T)\equiv\underline{\Theta}\Psi(b,T).
\ee
An initial solution at time $T_0$, satisfies the evolution equation
\be
\Psi(b,T)=e^{i\underline{\Theta}(T-T_0)}\Psi(b,T_0).
\ee
The space of solutions to (\ref{WDW}) can be equipped with the physical inner product
\be
(\Psi_1,\Psi_2)_{\text{phys}}=\int_{-\infty}^\infty db\,\bar{\Psi}_1(b,T_0)\Psi_2(b,T_0),
\ee
which is independent of the fixed instant of time $T_0$ because the evolution operator $\underline{\Theta}$ is self-adjoint with respect to it. Finally, we call the resulting physical Hilbert space $\underline{\mathfrak{H}}^{\text{phys}}$.

Now let us introduce the set of Dirac observables. The first is the cosmological constant $\hat{\Lambda}=-i\kappa\hbar\partial/\partial T$. It is a constant of motion. The second is the volume at any fixed instant of time, which gives a one-parameter family of relational observables defined as
\be\label{V value WDW}
\hat{V}|_{T_0}\Psi(b,T)=2\pi\gamma\lp^2e^{i\underline{\Theta}(T-T_0)}|\hat{\nu}|\Psi(b,T_0).
\ee
In the $b$ representation, the operator $\hat{\nu}$ acts like $-2i\partial/\partial b$. For any $|\Psi)\in\underline{\mathfrak{H}}^{\text{phys}}$, we have
\ba
(\Psi,\hat{V}|_{T_0}\Psi)_{\text{phys}}
&\equiv&\langle\hat{V}|_{T_0}\rangle\cr\cr
&=&4i\pi\gamma\lp^2\int_{-\infty}^\infty db\,\bar{\Psi}(b,T_0)\frac{\partial}{\partial b}\Psi(b,T_0)\cr\cr
&=&4i\pi\gamma\lp^2\int_{-\infty}^\infty db\,\bar{\Psi}(b,T_0)\frac{\partial}{\partial b}\bigg[e^{-i\underline{\Theta}(T-T_0)}\Psi(b,T)\bigg]\cr\cr
&=&(\Psi,\hat{V}|_T\Psi)_{\text{phys}}+\frac{3}{\gamma}(T-T_0)\int_{-\infty}^\infty db\,\bar{\Psi}(b,T_0)\,b\Psi(b,T_0).
\ea
Meanwhile, the expectation value for the cosmological constant is given by
\ba\label{Lambda value WDW}
(\Psi,\hat{\Lambda}\Psi)_{\text{phys}}\equiv\langle\hat{\Lambda}\rangle
&=&-i\kappa\hbar\int_{-\infty}^\infty db\,\bar{\Psi}(b,T)\frac{\partial\Psi(b,T)}{\partial T}\cr\cr
&=&\frac{3}{\gamma^2}\int_{-\infty}^\infty db\,\bar{\Psi}(b,T)\,b^2\Psi(b,T),
\ea
which is independent of $T$ as $\hat{\Lambda}$ commutes with $\underline{\Theta}$. Also note that (\ref{Lambda value WDW}) automatically yields $\langle\hat{\Lambda}\rangle>0$ (see footnote \ref{footnote on ADS}). Furthermore, we also define $\sqrt{\hat{\Lambda}}$ as the square root of $\hat{\Lambda}$ via its spectral decomposition, such that $\sqrt{\hat{\Lambda}}\,\sqrt{\hat{\Lambda}}=\hat{\Lambda}$. The matrix element of $\sqrt{\hat{\Lambda}}$ is thus given by
\be\label{sqrt Lambda value WDW}
(\Psi,\sqrt{\hat{\Lambda}}\Psi)_{\text{phys}}
\equiv\langle\sqrt{\hat{\Lambda}}\,\rangle
=\frac{\sqrt{3}}{\gamma}\int_{-\infty}^\infty db\,\bar{\Psi}(b,T)\,b\Psi(b,T).
\ee
It should be noted that $\sqrt{\hat{\Lambda}}\neq\sqrt{|\hat{\Lambda}|}$, and $\langle\sqrt{\hat{\Lambda}}\,\rangle$ can be positive or negative depending on the state $|\Psi)$.

Equations (\ref{V value WDW}) and (\ref{sqrt Lambda value WDW}) lead to
\be
\langle\hat{V}|_T\rangle-\langle\hat{V}|_{T_0}\rangle
=\sqrt{3}\,\langle\sqrt{\hat{\Lambda}}\,\rangle(T-T_0).
\ee
For a semiclassical quantum state which is sharply peaked (highly coherent), we have the good ``factorization approximation'' $\langle\hat{\Lambda}\rangle\approx\langle\sqrt{\hat{\Lambda}}\,\rangle^2$ or $\langle\sqrt{\hat{\Lambda}}\,\rangle\approx\pm\sqrt{\langle\hat{\Lambda}\,\rangle}$, and thus
\be
\langle\hat{V}|_T\rangle-\langle\hat{V}|_{T_0}\rangle
\approx\pm\sqrt{3\langle\hat{\Lambda}\rangle}\,(T-T_0).
\ee
Therefore, the quantum evolution of the WDW theory closely follows the classical trajectory described by (\ref{class}) if the state is sharply peaked. However, if the quantum state is not sharply peaked, the evolution can be considerably deviated from the classical trajectory (\textit{i.e.} the backreaction is appreciable).

\subsection{Loop quantum cosmology}
Now we are ready to investigate the properties of the LQC model with the unimodular clock variable. Elements $\Psi(\nu,T)\in\mathfrak{H}^{\text{kin}}=\mathfrak{H}^{\text{kin}}_{\text{grav}}\otimes\mathfrak{H}^{\text{kin}}_{\Lambda}$ are wave functions with finite kinematical norm
\be
||\Psi||^2=\int_{\mathbb{R}} dT \sum_\nu|\Psi(\nu,T)|^2,
\ee
and an orthonormal basis in $\mathfrak{H}^{\text{kin}}$ is given by $|\nu,T\rangle$ with
\be
\langle\nu,T|\nu',T'\rangle=\delta_{\nu\nu'}\delta(T,T').
\ee
Physical states satisfy the evolution equation (\ref{LQC}), and the physical scalar product is given by
\be\label{physicalSP}
(\Psi_1,\Psi_2)_{\text{phys}}=\frac{\lambda}{\pi}\sum_{\nu=4n\lambda}\bar{\Psi}_1(\nu,T_0)\Psi_2(\nu,T_0),
\ee
which is again independent of the fixed instant of time $T_0$ for the reasons mentioned above\footnote{Unlike the physical inner product defined in (\ref{physicalSP with phi}), which excludes the state with support at $\nu=0$, according to (\ref{physicalSP}), neither the state with support at $\nu=0$ nor that with the support at $\nu=\infty$ is excluded from the physical Hilbert space. This again has nothing to do with the issue of singularity resolution.}. A complete set of Dirac observables is again given by $\hat{\Lambda}$ and $\hat{V}|_{T_0}$. These two operators preserve the solutions to (\ref{LQC}) and are self-adjoint with respect to (\ref{physicalSP}) because of the self-adjointness of $\tilde{\Theta}$.

Now let us introduce more rigorously the $b$ representation \cite{ashtekar-c-s,chiou}. In the $b$ representation, the state $|\Psi\rangle$ gives the wave functional
\be
\Psi(b,T)\equiv\langle b|\Psi\rangle=\sum_{\nu=4n\lambda}\langle b|\nu\rangle\langle\nu|\Psi\rangle=\sum_{\nu=4n\lambda}e^{i\nu b/2}\Psi(\nu,T).
\ee
as the Fourier transform of $\Psi(\nu,T)$. From this we see that $\Psi(b+\pi/\lambda,T)=\Psi(b,T)$ is periodic, and its inverse Fourier transform is
\be
\Psi(\nu,T)\equiv\langle \nu|\Psi\rangle=\frac{\lambda}{\pi}\int^{\pi/\lambda}_0db\,\langle\nu|b\rangle\langle b|\Psi\rangle
=\frac{\lambda}{\pi}\int^{\pi/\lambda}_0db\,e^{-i\nu b/2}\Psi(b,T).
\ee
The symmetry requirement $\Psi(\nu,T)=\Psi(-\nu,T)$ that we have chosen to handle the gauge transformation $\Pi$ is here equivalent to $\Psi(b,T)=\Psi(-b,T)$. In the $b$ representation, the operator $\hat{\nu}$ acts like
\be
\hat{\nu}\Psi(b,T)=-2i\frac{\partial}{\partial b}\Psi(b,T),
\ee
the operator $\sin(\lambda b)$ acts as a multiplicative operator, and the evolution equation (\ref{LQC}) becomes
\be
-i\frac{\partial\Psi(b,T)}{\partial T}=\frac{3}{\kappa\hbar\gamma^2}\frac{\sin^2(\lambda b)}{\lambda^2}\Psi(b,T)=\tilde{\Theta}\Psi(b,T).
\ee
To obtain the physical inner product (\ref{physicalSP}) in the $b$ representation, we have to use the formula for the Fourier transform of a Dirac comb:
\be
\frac{1}{Q}\sum_{k=-\infty}^\infty\delta\left(f-\frac{k}{Q}\right)=\sum_{n=-\infty}^\infty e^{-2in\pi fQ},
\ee
which leads to
\be
\sum_{\nu=4n\lambda}e^{-i\nu(b-b')/2}=\frac{\pi}{\lambda}\sum_{k=-\infty}^\infty\delta\left(b-b'-k\frac{\pi}{\lambda}\right)=\frac{\pi}{\lambda}\delta(b-b')
\ee
if $b,b'\in[0,\pi/\lambda)$. Finally, we obtain
\be
(\Psi_1,\Psi_2)_{\text{phys}}=\int_0^{\pi/\lambda}db\,\bar{\Psi}_1(b,T_0)\Psi_2(b,T_0).
\ee

The matrix elements of the Dirac observables are given by
\ba
(\Psi,\hat{V}|_{T_0}\Psi)_{\text{phys}}
%:
&=&4i\pi\gamma\lp^2\int^{\pi/\lambda}_0db\,\bar{\Psi}(b,T_0)\frac{\partial}{\partial b}\Psi(b,T_0)\\
&=&4i\pi\gamma\lp^2\int^{\pi/\lambda}_0db\,\bar{\Psi}(b,T_0)\frac{\partial}{\partial b}\bigg[e^{-i\tilde{\Theta}(T-T_0)}\Psi(b,T)\bigg]\cr\cr
&=&(\Psi,\hat{V}|_T\Psi)_{\text{phys}}+\frac{3}{\gamma}(T-T_0)\int^{\pi/\lambda}_0db\,\bar{\Psi}(b,T_0)\frac{\sin(\lambda b)}{\lambda}\cos(\lambda b)\Psi(b,T_0),\nonumber
\ea
and
\ba
(\Psi,\hat{\Lambda}\Psi)_{\text{phys}}&=&-i\kappa\hbar\int^{\pi/\lambda}_0db\,\bar{\Psi}(b,T)\frac{\partial\Psi(b,T)}{\partial T}\cr\cr
&=&\frac{3}{\gamma^2}\int^{\pi/\lambda}_0db\,\bar{\Psi}(b,T)\frac{\sin^2(\lambda b)}{\lambda^2}\Psi(b,T),
\ea
as well as
\be
(\Psi,\sqrt{\hat{\Lambda}}\,\Psi)_{\text{phys}}
=\frac{\sqrt{3}}{\gamma}\int^{\pi/\lambda}_0db\,\bar{\Psi}(b,T)\frac{\sin(\lambda b)}{\lambda}\Psi(b,T).
\ee
Note that, as in the WDW theory, $\langle\hat{\Lambda}\rangle>0$.

In terms of the expectation values, the evolution is of the type
\be
\langle\hat{V}|_T\rangle-\langle\hat{V}|_{T_0}\rangle
=\sqrt{3}\left\langle\sqrt{\hat{\Lambda}}\,
\sqrt{1-\frac{\gamma^2\lambda^2}{6}\hat{\Lambda}}\right\rangle(T-T_0).
\ee
Again, if the state is sharply peaked, we can make the following factorization approximation:
\be
\langle\hat{V}|_T\rangle-\langle\hat{V}|_{T_0}\rangle
\approx\pm\sqrt{3\langle\hat{\Lambda}\rangle}\
\sqrt{1-\frac{\gamma^2\lambda^2}{6}\langle\hat{\Lambda}\rangle}\ (T-T_0).
\ee
Compared with the classical counterpart (\ref{class}), this shows that the expectation value of the cosmological constant has to be small ($\langle\hat{\Lambda}\rangle\ll\gamma^2\lambda^2$) to avoid strong departure from the classical trajectory. In this sense, the requirement of semiclassicality demands a small value of the cosmological constant. This is in agreement with the result obtained for the vertex expansion in the spin foam formulation of LQC as discussed in \cite{ACH2}.

\section{Nontrivial matter sector}\label{section7}
In this section we consider the quantization of the unimodular theory with generic matter sources. In the simplified LQC scheme and the $\nu$ representation, the quantum constraint equation (\ref{full}) takes the form
\be\label{fullquantum}
-i\frac{\partial\Psi(\nu,T)}{\partial T}=\left(\frac{3}{\kappa\hbar\gamma^2}\frac{\widehat{\sin(\lambda b)}}{\lambda}\frac{\widehat{\sin(\lambda b)}}{\lambda}-C_w|\hat{\nu}|^{-(w+1)}
-D\hat{p}^2_\phi|\hat{\nu}|^{-2}-\frac{1}{\hbar}U(\hat{\phi})\right)\Psi(\nu,T),
\ee
where
\be
C_w=\frac{\beta}{\hbar(2\pi\gamma\lp^2)^{w+1}},~~~~~~~~~~~~~~~D=\frac{1}{8\hbar\pi^2\gamma^2\lp^4}.
\ee

As our purpose of including matters is to create the classical singularity and see if it is resolved at the quantum level, in the following, we will consider the simple case by turning off the scalar field and including only one phenomenological matter.

\subsection{Wheeler-DeWitt limit}
In the formal limit $\lambda\rightarrow0$, equation (\ref{fullquantum}) with the scalar field turned off becomes
\be\label{matterWDW}
-i\frac{\partial\Psi(\nu,T)}{\partial T}=\left(\frac{3}{\kappa\hbar\gamma^2}\,\hat{b}^2-C_{w}|\hat{\nu}|^{-(w+1)}\right)\Psi(\nu,T)\equiv\underline{\Theta}\Psi(\nu,T).
\ee
In the $\nu$ representation, matrix elements of the volume operator are given by
\ba
(\Psi,\hat{V}|_{T_0}\Psi)_{\text{phys}}
&=&2\pi\gamma\lp^2\int_{-\infty}^\infty d\nu\,\bar{\Psi}(\nu,T_0)|\hat{\nu}|\Psi(\nu,T_0)\cr\cr
&=&2\pi\gamma\lp^2\int_{-\infty}^\infty d\nu\,\bar{\Psi}(\nu,T)e^{i\underline{\Theta}(T-T_0)}|\hat{\nu}|e^{-i\underline{\Theta}(T-T_0)}\Psi(\nu,T).
\ea
If we use the Baker-Campbell-Hausdorff formula to write
\be
e^{i\underline{\Theta}(T-T_0)}|\hat{\nu}|e^{-i\underline{\Theta}(T-T_0)}=|\hat{\nu}|+i(T-T_0)\big[\underline{\Theta},|\hat{\nu}|\big]-
\frac{1}{2}(T-T_0)^2\big[\underline{\Theta},\big[\underline{\Theta},|\hat{\nu}|\big]\big]+\dots,
\ee
the development will stop at order $n+1$ if the exponent of the operator $|\hat{\nu}|$ in (\ref{matterWDW}) is equal to $n\in\mathbb{N}$ (\textit{i.e.} if $-w\in\mathbb{N}^*$). For the sake of convenience, let us choose an exotic form of matter with $w=-2$, and define $C_{(w=-2)}\equiv C=\kappa\beta\gamma/4$. Then we find
\be
(\Psi,\hat{V}|_{T_0}\Psi)_{\text{phys}}
=(\Psi,\hat{V}|_T\Psi)_{\text{phys}}-\frac{3}{\gamma}(T-T_0)\int_{-\infty}^\infty d\nu\,\bar{\Psi}(\nu,T)\hat{b}\Psi(\nu,T)+\frac{3}{\gamma}C(T-T_0)^2,
\ee
and for the cosmological constant we get
\ba
(\Psi,\hat{\Lambda}\Psi)_{\text{phys}}&=&-i\kappa\hbar\int_{-\infty}^\infty d\nu\,\bar{\Psi}(\nu,T)\frac{\partial\Psi(\nu,T)}{\partial T}\cr\cr
&=&\frac{3}{\gamma^2}\int_{-\infty}^\infty d\nu\,\bar{\Psi}(\nu,T)\hat{b}^2\Psi(\nu,T)-\hbar\kappa\,C\int_{-\infty}^\infty d\nu\,\bar{\Psi}(\nu,T)|\hat{\nu}|\Psi(\nu,T).
\ea
With the inclusion of matter, $\langle\hat{\Lambda}\rangle$ can be positive or negative, depending on the quantum state.
The equations above lead to the evolution
\begin{subequations}\label{WDW with matter}
\ba
\langle\hat{V}|_T\rangle-\langle\hat{V}|_{T_0}\rangle
&=&\sqrt{3}\left\langle\sqrt{\hat{\Lambda}+\beta\kappa\hat{V}|_T}\,\right\rangle(T-T_0)
-\frac{3\beta\kappa}{4}(T-T_0)^2\\
&\approx&\pm\sqrt{3}\sqrt{\left\langle\hat{\Lambda}
+\beta\kappa\hat{V}|_T\right\rangle}\ (T-T_0)
-\frac{3\beta\kappa}{4}(T-T_0)^2,
\ea
\end{subequations}
where the square root of an operator is again defined through its spectral decomposition, and the factorization approximation has been used to obtain the second line. The evolution described in (\ref{WDW with matter}) closely follows the classical trajectory described by (\ref{class with matter}) if the quantum state is sharply peaked, and thus the big rip singularity is not avoided in the WDW theory.

\subsection{Loop quantum cosmology}
Without the scalar field, the LQC equation (\ref{fullquantum}) reads as
\be
-i\frac{\partial\Psi(\nu,T)}{\partial T}=\left(\frac{3}{\kappa\hbar\gamma^2}\frac{\widehat{\sin(\lambda b)}}{\lambda}\frac{\widehat{\sin(\lambda b)}}{\lambda}-C_w|\hat{\nu}|^{-(w+1)}
\right)\Psi(\nu,T)\equiv\tilde{\Theta}\Psi(\nu,T),
\ee
and the physical scalar product between two states is given by (\ref{physicalSP}). For $w=-2$, matrix elements of the Dirac observables are given by
\ba
(\Psi,\hat{V}|_{T_0}\Psi)_{\text{phys}}
&=&2\pi\gamma\lp^2\sum_{\nu=4n\lambda}\bar{\Psi}(\nu,T_0)|\hat{\nu}|\Psi(\nu,T_0)\cr\cr
&=&2\pi\gamma\lp^2\sum_{\nu=4n\lambda}\bar{\Psi}(\nu,T)e^{i\tilde{\Theta}(T-T_0)}|\hat{\nu}|e^{-i\tilde{\Theta}(T-T_0)}\Psi(\nu,T)\cr\cr
&=&(\Psi,\hat{V}|_T\Psi)_{\text{phys}}+\frac{3}{\gamma}(T-T_0)\sum_{\nu=4n\lambda}\bar{\Psi}(\nu,T)\widehat{\frac{\sin(\lambda b)}{\lambda}}\widehat{\cos(\lambda b)}\Psi(\nu,T)\cr\cr
&&-\frac{3}{\gamma}C(T-T_0)^2\sum_{\nu=4n\lambda}\bar{\Psi}(\nu,T)\left[1-2\widehat{\sin(\lambda b)}\widehat{\sin(\lambda b)}\right]\Psi(\nu,T),
\ea
and
\ba
(\Psi,\hat{\Lambda}\Psi)_{\text{phys}}&=&-i\kappa\hbar\sum_{\nu=4n\lambda}\bar{\Psi}(\nu,T)\frac{\partial\Psi(\nu,T)}{\partial T}\\
&=&\frac{3}{\gamma^2}\sum_{\nu=4n\lambda}\bar{\Psi}(\nu,T)\frac{\widehat{\sin(\lambda b)}}{\lambda}\frac{\widehat{\sin(\lambda b)}}{\lambda}\Psi(\nu,T)
-\hbar\kappa\, C\sum_{\nu=4n\lambda}\bar{\Psi}(\nu,T)|\hat{\nu}|\Psi(\nu,T).\nonumber
\ea
Again, $\langle\hat{\Lambda}\rangle$ can be positive or negative.

From these two equations we get
\begin{subequations}
\ba
\label{LQC with matter a}
\langle\hat{V}|_T\rangle-\langle\hat{V}|_{T_0}\rangle
&=&\sqrt{3}\left\langle\sqrt{\hat{\Lambda}+\beta\kappa\hat{V}|_T}\
\sqrt{1-\frac{\gamma^2\lambda^2}{3}
\left(\hat{\Lambda}+\beta\kappa\hat{V}|_T\right)}\right\rangle(T-T_0)\cr\cr
&&\quad-\frac{3\beta\kappa}{4}
\left(1-\frac{\gamma^2\lambda^2}{3}
\left\langle\hat{\Lambda}+\beta\kappa\hat{V}|_T\right\rangle\right)(T-T_0)^2\\
\label{LQC with matter b}
&\approx&\pm\sqrt{3\kappa}\sqrt{\left\langle\hat{\rho}_{\Lambda+w}|_T\right\rangle}\,
\left\langle\sqrt{1-\frac{\hat{\rho}_{\Lambda+w}|_T}{\rho_\text{crit}}}\,
\right\rangle(T-T_0)\cr\cr
&&\quad-\frac{3\beta\kappa}{4}
\left(1-\frac{\left\langle\hat{\rho}_{\Lambda+w}|_T\right\rangle}{\rho_\text{crit}}\right)
(T-T_0)^2,
\ea
\end{subequations}
where the total energy density operator at the instant $T$ is defined as
\be
\hat{\rho}_{\Lambda+w}|_T\equiv\frac{\hat{\Lambda}}{\kappa}+\beta\hat{V}|_T
\ee
in accordance with the classical counterpart (\ref{total density}), the critical density is given by
\be\label{crit density}
\rho_\text{crit}\equiv\frac{3}{\kappa\gamma^2\lambda^2},
\ee
and the square root of an operator is defined through its spectral decomposition. For the quantum state which is sharply peaked, the quantum evolution closely follows the classical trajectory (\ref{class}) in the classical regime when $\left\langle\hat{\rho}_{\Lambda+w}|_T\right\rangle\ll\rho_\text{crit}$. Towards the (putative) big rip singularity, $\langle\hat{V}|_T\rangle$ keeps growing and so does $\left\langle\hat{\rho}_{\Lambda+w}|_T\right\rangle$, taking the evolution away from the classical regime. Eventually, as $\left\langle\hat{\rho}_{\Lambda+w}|_T\right\rangle$ approaches $\rho_\text{crit}$, the factor
\be
\left\langle\sqrt{1-\frac{\hat{\rho}_{\Lambda+w}|_T}{\rho_\text{crit}}}\,\right\rangle \approx \pm\sqrt{1-\frac{\left\langle\hat{\rho}_{\Lambda+w}|_T\right\rangle}{\rho_\text{crit}}}
\ee
in (\ref{LQC with matter b}) flips signs and $\langle\hat{V}|_T\rangle$ gets bounced and starts to decrease. Therefore, the expectation value of the total density is prevented from blowing up by the quantum bounce. By the same notion of singularity resolution suggested in \cite{ashtekar-c-s}, the classical big rip singularity is said to be resolved and replaced by the quantum bounce, which bridges the expanding universe with the contracting one. The quantum bounce takes place at the epoch when $\left\langle\hat{\rho}_{\Lambda+w}|_T\right\rangle \approx \rho_\text{crit}$. The critical density $\rho_\text{crit}$ given in (\ref{crit density}) is exactly the same as that for other models of LQC \cite{aps,ashtekar-c-s}. The fact that the occurrence of the quantum bounce is signaled by the total (matter plus cosmological constant) energy density  also agrees with the LQC model featuring a scalar field \cite{aps}.

\section{Spin foam formulation}\label{section8}
Spin foam models for LQG were introduced as a sum over histories of quantum spin network states in a seminal work by Reisenberger and Rovelli \cite{RR}. Interestingly, many different avenues to quantum gravity have converged to spin foam models, which now constitute a very active research area \cite{rovelli-book,thiemann-book}. Despite the numerous deep technical results that have been obtained over the years, several open issues still remain, and it has recently been argued that LQC can be used to shed light on these problems \cite{ACH2,ACH1}. In particular, it has been realized that the Hamiltonian theory lying behind LQC can be written as a sum over histories that renders the analogy with spin foam models more transparent. Let us first briefly recall the ideas behind this formulation and the key results that have been obtained so far. Then we will see how the unimodular time coordinate can be used to clarify some issues stemming from the standard LQC vertex expansion.

\subsection{Vertex expansion with a scalar field clock}
In LQC with a scalar field, there are two distinct but equivalent ways to look at the quantum constraint equation (\ref{LQCev}) and construct the physical state space. The first one is to consider the scalar field as an internal clock and then construct the positive frequency solutions to the Klein-Gordon equation (\ref{sqrtLQC}). In this deparametrized framework, the main object of interest is the transition amplitude
\be
A(\nu_f,\phi_f;\nu_i,\phi_i)=\langle\nu_f|e^{i\sqrt{\Theta}(\phi_f-\phi_i)}|\nu_i\rangle
\ee
for an initial physical state $|\nu_i\rangle$ at internal time $\phi_i$ to evolve to a final state $|\nu_f\rangle$ at internal time $\phi_f$.
The alternative path is to consider the quantum constraint $\mathcal{C}=-\partial^2_\phi-\Theta$ as a whole and use the group averaging procedure \cite{marolf1,marolf2,marolf3} to map the kinematical states into the kernel of the constraint operator. This is more in the spirit of the spin foam approach, where one does not have a preferred internal time variable to deparametrize the theory. In this so-called timeless framework, attention is focused on kinematical states $|\nu_i,\phi_i\rangle$ and $|\nu_f,\phi_f\rangle$ in $\mathfrak{H}^{\text{kin}}$, and the objects of interest are the matrix elements of the group averaged inner product, given by the Green functions
\be
G(\nu_f,\phi_f;\nu_i,\phi_i)=2\int d\alpha\,\langle\nu_f,\phi_f|e^{i\alpha\mathcal{C}}|p_\phi||\nu_i,\phi_i\rangle.
\ee
These two ways of constructing the physical Hilbert space are equivalent, and it has been shown in \cite{ACH2} that one has the equality $G(\nu_f,\phi_f;\nu_i,\phi_i)=A(\nu_f,\phi_f;\nu_i,\phi_i)$. The idea of the Ashtekar-Campiglia-Henderson expansion is then to use a sum over histories formulation to obtain a series expansion for these two quantities which reproduces in spirit the vertex expansion of spin foam models.

The vertex expansions that have been obtained \cite{ACH2} for the transition amplitude and the physical inner product are respectively given by
\ba
A(\nu_f,\phi_f;\nu_i,\phi_i)
&=&\sum_{M=0}^\infty\left[\sum_{\substack{\nu_{M-1},\dots,\nu_1\\\nu_m\neq\nu_{m+1}}}
H_{\nu_M\nu_{M-1}}H_{\nu_{M-1}\nu_{M-2}}\dots H_{\nu_1\nu_0}\times\right.\\
&&\left.\prod_{k=1}^p\frac{1}{(n_k-1)!}\left(\frac{\partial}{\partial H_{w_kw_k}}\right)^{n_k-1}
\sum_{m=1}^p\frac{e^{iH_{w_mw_m}(\phi_f-\phi_i)}}{\prod_{j\neq m}^p(H_{w_mw_m}-H_{w_jw_j})}\right],\nonumber
\ea
and
\ba
G_+(\nu_f,\phi_f;\nu_i,\phi_i)&=&\sum_{M=0}^\infty\left[\sum_{\substack{\nu_{M-1},\dots,\nu_1\\\nu_m\neq\nu_{m+1}}}
\Theta_{\nu_M\nu_{M-1}}\Theta_{\nu_{M-1}\nu_{M-2}}\dots\Theta_{\nu_1\nu_0}\times\right.\label{SFexp}\\
&&\left.\prod_{k=1}^p\frac{1}{(n_k-1)!}\left(\frac{\partial}{\partial\Theta_{w_kw_k}}\right)^{n_k-1}
\sum_{m=1}^p\frac{e^{i\sqrt{\Theta_{w_mw_m}}(\phi_f-\phi_i)}}{\prod_{j\neq m}^p(\Theta_{w_mw_m}-\Theta_{w_jw_j})}\right],\nonumber
\ea
where $H\equiv\sqrt{\Theta}$, and the subscript $+$ refers to the positive frequency part of the group averaged kinematical states. These expressions of the transition amplitude and the inner product are in the form of a discrete sum over paths containing $M$ volume transitions. A key observation is that although the objects that were computed are the same, they have two distinct vertex expansions. One is indeed featuring matrix elements of the operator $\sqrt{\Theta}$ whereas the other uses matrix elements of $\Theta$. This peculiarity can be traced back to the form of the Hamiltonian constraint of LQC and the fact that the clock variable that is used for deparametrization is introduced through $p_\phi^2$. In unimodular theory, the fact that the Hamiltonian constraint is linear in the momentum conjugate to the time variable enables us to circumvent this difficulty. To see this, we will focus on the timeless framework and reproduce the main steps of the procedure leading to the sum over histories.

\subsection{Vertex expansion with the unimodular time variable}
In the spin foam approach, one does not have a preferred time variable which serves as a relational time, and the emphasis is made on the timeless formalism. We will start here with the full constraint operator
\be\label{constraint}
\mathcal{C}\equiv\tilde{\Theta}+i\frac{\partial}{\partial T}=\tilde{\Theta}-\Lambda=0,
\ee
which is self-adjoint on the kinematical Hilbert space $\mathfrak{H}^\text{kin}$, and use the group averaging procedure to construct the physical inner product. The procedure is the following. We begin by fixing a dense subspace $\mathcal{S}\subset\mathfrak{H}^\text{kin}$, which can be taken to be the Schwartz space of rapidly decreasing functions $f(\nu,T)$. If we denote by $e_k(\nu),\;k\in(-\infty,\infty)$, orthogonal eigenfunctions of $\tilde{\Theta}$ with eigenvalue $\omega_k$, any $f(\nu,T)\in\mathcal{S}$ can be expanded as
\be
f(\nu,T)=\frac{1}{2\pi}\int dk\int d\Lambda\,\tilde{f}(k,\Lambda)e^{i\Lambda T}e_k(\nu).
\ee
Distributional solutions to the constraint equation lie in the topological dual $\mathcal{S}^\star$ of $\mathcal{S}$, and the inclusion relation between all the spaces involved is $\mathcal{S}\subset\mathfrak{H}^\text{kin}\subset\mathcal{S}^\star$. The group averaging procedure \cite{marolf1,marolf2,marolf3} tells us that the solutions are explicitly given by
\ba
\Psi_f(\nu,T)&=&\int d\alpha\,e^{i\alpha\mathcal{C}}f(\nu,T)\cr\cr
&=&\int dk\int d\Lambda\,\delta(\Lambda-\omega_k)\tilde{f}(k,\Lambda)e^{i\Lambda T}e_k(\nu)\cr\cr
&=&\int dk\,\tilde{f}(k,\omega_k)e^{i\omega_kT}e_k(\nu),
\ea
and they satisfy the evolution equation
\be
\Psi_f(\nu,T)=e^{i\tilde{\Theta}(T-T_0)}\Psi_f(\nu,T_0).
\ee
Given a distribution $\Psi_f\in\mathcal{S}^\star$ associated with $f\in\mathcal{S}$, its linear action on an arbitrary function $g\in\mathcal{S}$ is given by
\ba
(\Psi_f|g\rangle
&=&\int dk\int d\Lambda\,\delta(\Lambda-\omega_k)\bar{\tilde{f}}(k,\Lambda)\tilde{g}(k,\Lambda)\cr\cr
&=&\int dk\,\bar{\tilde{f}}(k,\omega_k)\,\tilde{g}(k,\omega_k).
\ea
Now given two elements $f,g\in\mathcal{S}$, the scalar product between the corresponding states in $\mathcal{S}^\star$ can be defined as
\be
(\Psi_f,\Psi_g)_{\text{phys}}\equiv(\Psi_f|g\rangle=\overline{(\Psi_g|f\rangle}.
\ee
If we consider the full constraint (\ref{constraint}), the object of interest is this physical inner product. Instead of considering general kinematical states $f(\nu,T)$, we are going to focus on the basis of $\mathfrak{H}^{\text{kin}}$.

Let us choose two states $|\nu_i,T_i\rangle$ and $|\nu_f,T_f\rangle$ in $\mathfrak{H}^{\text{kin}}$. The group averaged inner product between these two states is given by
\be\label{green}
G(\nu_f,T_f;\nu_i,T_i)=\int d\alpha\,\langle\nu_t,T_f|e^{i\alpha\mathcal{C}}|\nu_i,T_i\rangle=\int d\alpha\,A(\nu_f,T_f;\nu_i,T_i;\alpha),
\ee
where $A$ is the probability amplitude for a kinematical state $|\nu_i,T_i\rangle$ to ``evolve'' to another state $|\nu_f,T_f\rangle$ through the unitary transformation by $\exp(i\alpha\,\mathcal{C})$. Since the full constraint $\mathcal{C}$ is the sum of two commuting operators acting separately on $\mathfrak{H}^{\text{kin}}_{\text{grav}}$ and $\mathfrak{H}^{\text{kin}}_{\Lambda}$, we can write the amplitude $A$ as
\be
A(\nu_f,T_f;\nu_i,T_i;\alpha)=A_{\text{grav}}(\nu_f;\nu_i;\alpha)A_T(T_f;T_i;\alpha),
\ee
where
\be
A_{\text{grav}}(\nu_f;\nu_i;\alpha)=\langle\nu_f|e^{-i\alpha\tilde{\Theta}}|\nu_i\rangle,
\ee
and
\be
A_T(T_f;T_i;\alpha)=\langle T_f|e^{i\alpha\Lambda}|T_i\rangle=\int d\Lambda\,e^{i\alpha\Lambda}e^{i\Lambda(T_f-T_i)}.
\ee
Now we need to write down the expansion of the gravitational amplitude. To do so, we rigorously follow the treatment carried out in \cite{ACH2}, and express it as a sum over discrete histories. It can be shown that the amplitude takes the form
\ba
A_{\text{grav}}(\nu_f;\nu_i;\alpha)
&=&\sum_{M=0}^{+\infty}\left[\sum_{\substack{\nu_{M-1},\dots\nu_{1}\\\nu_m\neq\nu_{m+1}}}
\tilde{\Theta}_{\nu_M\nu_{M-1}}\tilde{\Theta}_{\nu_{M-1}\nu_{M-2}}\dots\tilde{\Theta}_{\nu_1\nu_0}\times\right.\\
&&\left.\prod_{k=1}^p\frac{1}{(n_k-1)!}\left(\frac{\partial}{\partial\tilde{\Theta}_{w_kw_k}}\right)^{n_k-1}\sum_{m=1}^p
\frac{e^{-i\alpha\tilde{\Theta}_{w_mw_m}\Delta t}}{\prod_{j\neq m}^p\big(\tilde{\Theta}_{w_mw_m}-\tilde{\Theta}_{w_jw_j}\big)}\right],\nonumber
\ea
where $\Delta t=1$ is the time interval used for the skeletonization of the Feynman path integral. Now if we make the assumption that the sum over $M$ in this expression commutes with the integration over $\alpha$ in (\ref{green}), we find that the group averaged inner product takes the form
\ba\label{expansionG}
G(\nu_f,T_f;\nu_i,T_i)&=&\sum_{M=0}^{+\infty}\left[\sum_{\substack{\nu_{M-1},\dots\nu_{1}\\\nu_m\neq\nu_{m+1}}}
\tilde{\Theta}_{\nu_M\nu_{M-1}}\tilde{\Theta}_{\nu_{M-1}\nu_{M-2}}\dots\tilde{\Theta}_{\nu_1\nu_0}\times\right.\cr\cr
&&\left.\prod_{k=1}^p\frac{1}{(n_k-1)!}\left(\frac{\partial}{\partial\tilde{\Theta}_{w_kw_k}}\right)^{n_k-1}\sum_{m=1}^p
\int d\Lambda\,e^{i\Lambda(T_f-T_i)}\frac{\delta\big(\Lambda-\tilde{\Theta}_{w_mw_m}\Delta t\big)}{\prod_{j\neq m}^p\big(\tilde{\Theta}_{w_mw_m}-\tilde{\Theta}_{w_jw_j}\big)}\right]\cr\cr
&=&\sum_{M=0}^{+\infty}\left[\sum_{\substack{\nu_{M-1},\dots\nu_{1}\\\nu_m\neq\nu_{m+1}}}
\tilde{\Theta}_{\nu_M\nu_{M-1}}\tilde{\Theta}_{\nu_{M-1}\nu_{M-2}}\dots\tilde{\Theta}_{\nu_1\nu_0}\times\right.\cr\cr
&&\left.\prod_{k=1}^p\frac{1}{(n_k-1)!}\left(\frac{\partial}{\partial\tilde{\Theta}_{w_kw_k}}\right)^{n_k-1}\sum_{m=1}^p
\frac{e^{i\tilde{\Theta}_{w_mw_m}(T_f-T_i)}}{\prod_{j\neq m}^p\big(\tilde{\Theta}_{w_mw_m}-\tilde{\Theta}_{w_jw_j}\big)}\right].
\ea
This expression is to be compared with the expansion (\ref{SFexp}) that has been obtained with the matter clock.

If one rather focuses on the deparametrized framework and considers states satisfying the evolution equation
\be
-i\frac{\partial\Psi(\nu,T)}{\partial T}=\tilde{\Theta}\Psi(\nu,T),
\ee
it can be shown that the vertex expansion of the transition amplitude
\be
A(\nu_f,T_f;\nu_i,T_i)=\langle\nu_f|e^{i\tilde{\Theta}(T_f-T_i)}|\nu_i\rangle
\ee
is exactly the same as (\ref{expansionG}) at any order. Therefore not only do we have $A(\nu_f,T_f;\nu_i,T_i)=G(\nu_f,T_f;\nu_i,T_i)$, but now the two vertex expansions do coincide. This can be traced back to the fact that in the unimodular theory the time variable is introduced via a linear function of its conjugate momentum. Thus, there is one unique vertex expansion and its expression is simplified since there is no square root involved.

\section{Conclusion and discussion}\label{section9}

At the classical level, unimodular gravity is equivalent to Einstein's general relativity. The only difference resides in the fact that it allows for a dynamical cosmological constant. The price to pay for this particularity is that one has to extend the phase space in order to add the cosmological constant and its canonically conjugate momentum, which turns out to be a geometric time variable measuring the four volume between two hypersurfaces. The main aim of this paper was to investigate whether it is possible to use the cosmological reduction of unimodular theory as a starting point to build the quantum theory in the spirit of LQC. There is hope that this framework might be useful to address issues related to the problem of time and the cosmological constant \cite{smolin-cc}, and we think that it is important to study simplified models in order to clarify the statements and give an idea of how the construction of a quantum theory can be carried out.

It has been shown that when the cosmological time variable that arises in the unimodular theory as space-time four volume is used, the quantum theory appears to be simpler than that of usual LQC with a scalar field. By this, we mean that since the Hamiltonian is linear in the cosmological constant, which is conjugate to the unimodular time variable, the dynamical equation that is obtained by imposing the operatorial constraint at the quantum level is a Schr\"odinger-like equation. To investigate the relevance of this model, we have first studied its classical dynamics with a pure cosmological constant, and then with matter with a nontrivial equation of state in order to introduce a singularity. In the quantum theory, we have shown that the Wheeler-DeWitt dynamics does not lead to singularity resolution. The classical trajectory is followed and the (big rip) singularity is not avoided. In the LQC dynamics, however, the singularity is resolved and replaced by a quantum bounce that occurs when the expectation value of the total matter density reaches the critical value $\rho_{\text{crit}}=3/(\kappa\gamma^2\lambda^2)$. This critical value is in accordance with the one that has been previously derived in LQC with a scalar field clock \cite{aps,ashtekar-c-s}. Moreover, we have shown that in order for a classical solution to exist at late times in the framework of LQC, the expectation value of the cosmological constant has to be small enough (in Planck units), with an upper bound in agreement with what is usually expected from vacuum energy contributions in flat space-time quantum field theory. We see this result as a small clue indicating that unimodular theory might be useful to study the quantum dynamics of cosmological models. Indeed, it reproduces precisely the results that have been obtained previously, and leads to a linear evolution equation, which simplifies the study of the dynamics. We have also briefly studied the vertex expansion of LQC in the spirit of spin foam models, and shown that it is possible to obtain a unique expansion when the clock that is used is not a scalar field. It has been claimed \cite{ACH2} that different vertex expansions can appear whenever there is a degree of freedom in the theory (which has been taken to be a scalar field in previous studies) which we can use for deparametrization. In the unimodular theory, the timeless and deparametrized frameworks are trivially equivalent because of the linearity of the constraint in the momentum conjugate to the time variable. This supports the idea that two vertex expansions will exist if a matter clock is used for deparametrization, regardless of what happens in the gravitational sector (e.g. for Bianchi models).

There are several directions that one could follow from here. First of all, it is possible to use the unimodular theory to investigate more complicated cosmological models such as Bianchi models, Kasner models or the Schwarzschild interior \cite{boehmer,chiou-bh}. This might be a way to introduce a cosmological constant along with a time variable in order to compute the dynamics and the evolution of the variables of interest. Since the scalar field is not used as a clock anymore, it is now possible to quantize it in the polymer representation \cite{als}, and thereby obtain a full loop quantum description of the dynamics near and through the bounce. Alternatively, if one takes unimodular theory seriously, it is possible to further investigate its quantization in the loop representation. This approach might help to understand the regularization of the Hamiltonian constraint in the canonical approach, and the dynamics in the presence of a cosmological constant. It is indeed quite easy to cast the unimodular action of Henneaux and Teitelboim in the Plebanski formulation, which can be taken as a starting point for the spin foam quantization. Unfortunately, in four space-time dimensions, we do not have for the moment a good understanding of how to build the full quantum theory in the presence of a cosmological constant.

\begin{acknowledgements}
M.G. would like to thank Edward Anderson, Adam Henderson, Marc Lachi\`eze-Rey, Karim Noui, Lee Smolin and Edward Wilson-Ewing for helpful discussions and comments, as well as the Department of Physics at Beijing Normal University, where part of this work has been done. D.W.C. would like to acknowledge the support of Grant No. 10675019 from the NSFC and the support of Grants No. 20080440017 and No. 200902062 from the China Postdoctoral Science Foundation.
\end{acknowledgements}

\end{document}